# Newly synthesized $Zr_2AlC$, $Zr_2(Al_{0.58}Bi_{0.42})C$, $Zr_2(Al_{0.2}Sn_{0.8})C$, and $Zr_2(Al_{0.3}Sb_{0.7})C$ MAX phases: a DFT based first-principles study

M. A. Ali[a], M. M. Hossain[a], N. Jahan[a], A. K. M. A. Islam[b], S. H. Naqib[c*]

[a]Department of Physics, Chittagong University of Engineering and Technology, Chittagong-4349, Bangladesh.
[b]International Islamic University Chittagong, 154/A College Road, Chittagong, Bangladesh.
[c]Department of Physics, University of Rajshahi, Rajshahi-6205, Bangladesh.

**ABSTRACT**

The structural, elastic, and electronic properties of newly synthesized $Zr_2(Al_{0.58}Bi_{0.42})C$, $Zr_2(Al_{0.2}Sn_{0.8})C$, and $Zr_2(Al_{0.3}Sb_{0.7})C$ MAX nanolaminates have been studied using first-principles density functional theory (DFT) calculations for the first time. Theoretical Vickers hardness has also been estimated for these compounds. All the calculated results are compared with experimental data and also with that of recently discovered $Zr_2AlC$ phase, where available. $Zr_2(Al_{0.58}Bi_{0.42})C$ and $Zr_2(Al_{0.2}Sn_{0.8})C$ are the two first Bi and Sn containing MAX compounds. The calculated structural parameters are found to be in good agreement with the experimental data. The single crystal elastic constants $C_{ij}$ and other polycrystalline elastic coefficients have been calculated and the mechanical stabilities of these compounds have been theoretically confirmed. The bulk modulus increases and the shear modulus decreases due to partial Bi/Sn/Sb substitution for Al in $Zr_2AlC$. The calculated elastic moduli show that these Bi/Sn/Sb containing MAX phases are more anisotropic than $Zr_2AlC$, and have a tendency towards ductility. The Vickers hardness decreases in the Bi/Sn/Sb containing compounds. Further, the electronic band structures and electronic density of states (EDOS) are calculated and the effects of different elemental substitution on these properties are investigated. The electronic band structures show metallic characteristics with contribution predominantly coming from the Zr 4*d* orbitals. Partial presence of Bi/Sn/Sb atoms increases the EDOS at the Fermi level to some extent. Possible implications of the theoretical results for these recently discovered MAX nanolaminates have been discussed in detail in this paper.

*Corresponding author. Email: salehnaqib@yahoo.com

## 1. Introduction

A family of ternary carbides and nitrides, known to the scientific community as MAX phase with general formula $M_{n+1}AX_n$ where *n* is an integer, has attracted significant attention both experimentally and theoretically for their huge potential in variety of industrial applications. In $M_{n+1}AX_n$, M represents an early transition metal, A represents an A group element and X represents C and/or N atom [1]. Generally MAX phases crystallize with hexagonal structure with space group *P6₃/mmc*. The MAX phases exhibit unique properties combining the advantages of metals as well as ceramics due to their nanolaminated crystal structure with metallic layers containing A atoms in between the MX ceramic layers [1 – 3]. MAX compounds are resistant to corrosion and oxidation and they have high stiffness constant. Besides, these compounds have high tolerance to thermal shock, reasonable electrical conductivity, relatively high dielectric constant, and attractive optical properties for optoelectronic applications, and excellent machinability [3]. These remarkable set of properties have triggered noticeable activity within the scientific circle and study of MAX phases has become an important subfield in materials science and technology these days.

The MAX phases with $M_2AC$ stoichiometry was synthesized for the first time in the 1960s [2]. These compounds were almost ignored for two decades. Later on, in 1996, Barsoum and El-Raghy [4] reported some technologically important properties of the ceramic compound $Ti_3SiC_2$ to highlight the generic characteristics of MAX phases. Since then, similar properties have been found and reported for a large number of MAX compounds belonging to different families [3, 5, 6]. Hu *et* al. have reported a partial list of about 70 new MAX phases in the period 2004–2013 [7]. Recently, Naguib et al. [8] reported a large number of work (68 quaternary MAX phases) on the solid solutions where existing phases are mixed. These approaches yield the new phases [9, 10] and/or other ordered quaternary phases [8, 11 – 14]. Furthermore, the recent discovery of $Mo_2Ga_2C$ (with double "A" layers), possibly be the first of a distinct family of MAX-related phases adds to the diversity [15]. More recently, Lapauw *et al.* [16] have synthesized the compounds $Zr_2AlC$ and $Zr_3AlC_2$. The latter one was the first experimentally produced MAX phase in the Zr-Al-C system [16, 17]. More recently, theoretical works on the calculations of structural, mechanical, electronic, optical, vibrational and thermodynamics properties of some MAX phases such as $Ti_{n+1}GaN_n$ [18], $Sc_2AlC$ [19] and $Mo_2TiAlC_2$ [20] are also reported.

The ternary phase $Zr_2AlC$ is one of the most interesting among the MAX compounds. This is because it is expected to have high degree of neutron transparency because all the constituent atoms have small reaction cross-section for thermal neutrons, especially the Zr atom. It also has the potential to be used as a protective material in modified accident tolerant fuel [21]. Besides, this compound possesses the other generic features common to the MAX nanolaminates. In recent times Horlait *et al.* synthesized the first Bi containing solid solutions $Zr_2(Al_{1-x}Bi_x)C$ ($0 \leq x \leq 1$) [22] and also produced other solid solutions such as $Zr_2(Al_{0.2}Sn_{0.8})C$, $Zr_2(Al_{0.35}Pb_{0.65})C$, and $Zr_2(Al_{0.3}Sb_{0.7})C$ [21, 23]. Against this backdrop, very recent synthesis of partially substituted $Zr_2AlC$ compound, namely $Zr_2(Al_{0.58}Bi_{0.42})C$, $Zr_2(Al_{0.2}Sn_{0.8})C$ and $Zr_2(Al_{0.3}Sb_{0.7})C$, is an important milestone. To the best of our knowledge, no complete theoretical study on the structural, elastic, and electronic band structure of $Zr_2(Al_{0.58}Bi_{0.42})C$, $Zr_2(Al_{0.2}Sn_{0.8})C$ and $Zr_2(Al_{0.3}Sb_{0.7})C$ has been reported, so far. This background motivates us to investigate the properties of these newly synthesized ternary MAX phases by means of first principles calculations and to compare the results to those obtained for the $Zr_2AlC$ compound for the first time. In present investigation, the structural, elastic, and electronic band structures along with the theoretical Vickers hardness and bonding properties of these newly synthesized compounds are reported.

We organize the paper as follows. Section 2 consists of a brief description of the first principles computational methodology. All the theoretical results are shown and discussed in Section 3. Finally, the conclusions from this investigation are presented in Section 4.

## 2. Computational methodology

The first-principles calculations presented in this paper were performed using the Cambridge Serial Total Energy Package (CASTEP) code [24]. CASTEP employs the plane wave pseudopotential approach based on the density functional theory [25]. During the *ab-initio* calculations the exchange-correlation potential are treated within the generalized gradient approximation (GGA) using Perdew-Burke-Ernzerhof (PBE) functional [26]. This permits a variation in the electron density inside the crystal under study. Vanderbilt-type ultrasoft pseudopotentials were used to model the electron-ion interactions. A $10 \times 10 \times 2$ *k*-point mesh of Monkhorst-Pack scheme was used to sample the first Brillouin zone [27]. Throughout the computation, a plane-wave cutoff kinetic energy of 500 eV was chosen to determine the number of plane waves within the expansion. The crystal structures were fully relaxed by the widely employed Broyden-Fletcher-Goldfrab-Shanno (BFGS) optimization technique [28]. The various tolerances for the self-consistent field, energy, maximum force, maximum displacement, and maximum stress were set to be $5.0 \times 10^{-7}$ eV/atom, $5.0 \times 10^{-6}$ eV/atom, 0.01 eV/Å, $5.0 \times 10^{-4}$ Å, and 0.02 GPa, respectively.

## 3. Results and discussion

### 3.1 Structural properties

The ternary layered carbide $Zr_2AlC$ compound with space group *P63/mmc* belongs to the hexagonal system like other MAX phase compounds [3]. The unit cell contains two formula units. The atomic coordinates are as follows: Zr atoms are located at (1/3, 2/3, z), Al atoms at (2/3, 1/3, 1/4) and C atoms at (0, 0, 0). Bi/Sn/Sb atoms are partially substituted for Al at (2/3, 1/3, 1/4). Fig. 1 shows the unit cell of $Zr_2AlC$. The equilibrium crystal structure of this compound is obtained by minimizing the total energy. Optimum structural parameters are presented in Table 1 for pure and Bi/Sn/Sb containing $Zr_2AlC$ compounds. The results are in good agreement with the reported experimental values, where available. From Table 1, it is seen that the lattice constants of substituted compounds are not same as that of the $Zr_2AlC$. An increase in lattice constant $a$ is found while a decrease in $c$ is observed due to the Bi/Sn/Sb substitution for Al in $Zr_2AlC$. The change in lattice constants can be largely attributed to the ionic size differences between Bi/Sn/Sb and Al.

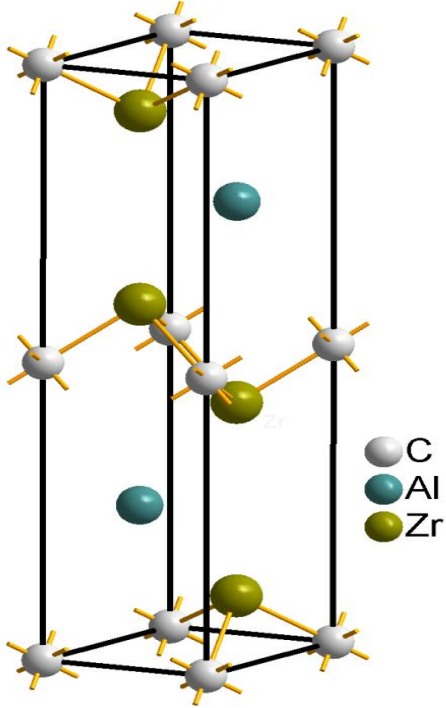

**Fig. 1.** Crystal structure of $Zr_2AlC$.

**Table 1**
Optimized lattice parameters ($a$ and $c$, in Å), hexagonal ratio $c/a$, internal parameter $z_M$, unit cell volume $V$ (Å$^3$) for $Zr_2(Al_{0.58}Bi_{0.42})C$, $Zr_2(Al_{0.2}Sn_{0.8})C$, $Zr_2(Al_{0.3}Sb_{0.7})C$, and $Zr_2AlC$ MAX phases.

| Phases | $a$ | $C$ | $c/a$ | $z_M$ | $V$ | Reference |
|---|---|---|---|---|---|---|
| $Zr_2(Al_{0.58}Bi_{0.42})C$ | 3.375 | 13.719 | 4.06 | 0.0923 | 135.372 | This study |
|  | 3.344 | 14.510 |  |  |  | Expt.[22] |
| $Zr_2(Al_{0.2}Sn_{0.8})C$ | 3.352 | 13.747 | 4.10 | 0.0922 | 133.796 | This study |
|  | 3.345 | 14.567 |  |  |  | Expt.[21] |
| $Zr_2(Al_{0.3}Sb_{0.7})C$ | 3.357 | 14.311 | 4.26 | 0.0885 | 139.699 | This study |
|  | 3.367 | 14.620 |  |  |  | Expt.[21] |
| $Zr_2AlC$ | 3.319 | 14.604 | 4.40 | 0.0864 | 139.300 | This study |
|  | 3.324 | 14.570 |  | 0.0871 |  | Expt.[23] |
|  | 3.319 | 14.606 | 4.40 | 0.0864 |  | Theo.[29] |

*Calculated using the relation $V = 0.866a^2c$

## 3.2 Elastic properties

The elastic constants give relationships connecting the dynamical and mechanical properties of crystalline materials. They are intimately related to the nature of electronic bonding among the atoms. Elastic constants determine the dynamical and mechanical responses of a compound under applied stress and are therefore, extremely important as far as engineering applications are concerned [30].

The elastic constants of $Zr_2(Al_{0.58}Bi_{0.42})C$, $Zr_2(Al_{0.2}Sn_{0.8})C$, $Zr_2(Al_{0.3}Sb_{0.7})C$ and $Zr_2AlC$ at ambient pressure have been calculated and are given in Table 2. The elastic constants have been calculated by computing the resulting stress generated due to applying a set of given homogeneous deformation of a finite value within the CASTEP code. Since the elastic constants for the substituted MAX compounds are estimated for the first time and no experimental study exists, we cannot compare them with any other result. A reasonable theoretical comparison can be made with the $Zr_2AlC$ phase, for which no study (theoretical or experimental) of elastic properties exists as well. A compound is considered to be mechanically stable if the single crystal elastic constants $C_{ij}$ are all positive and satisfy the well known Born criteria [31]: $C_{11} > 0$, $(C_{11} - C_{12}) > 0$, $C_{44} > 0$ and $(C_{11} + C_{12})C_{33} > 2C_{13}^2$. The compounds under consideration satisfy these criteria quite well as shown in Table 3 and are therefore, mechanically stable. From Table 2 it becomes obvious that due to the Bi/Sn/Sb substitution for Al in $Zr_2AlC$, most of the single crystal elastic constants increase significantly, except $C_{11}$ that represents the elasticity in length related to the longitudinal strain.

**Table 2**
The elastic constants $Cij$ (GPa), bulk modulus, $B$ (GPa), shear modulus, $G$ (GPa), Young's modulus, $Y$ (GPa), Poisson ratio, $v$, and Pugh ratio $G/B$, Vickers hardness $Hv$.

| Compound | $C_{11}$ | $C_{12}$ | $C_{13}$ | $C_{33}$ | $C_{44}$ | B | G | Y | G/B | v | Hv |
|---|---|---|---|---|---|---|---|---|---|---|---|
| $Zr_2(Al_{0.58}Bi_{0.42})C$ | 242 | 99 | 124 | 284 | 129 | 161 | 90 | 227 | 0.55 | 0.26 | 10.81 |
| $Zr_2(Al_{0.2}Sn_{0.8})C$ | 253 | 108 | 110 | 293 | 109 | 161 | 88 | 223 | 0.54 | 0.27 | 10.34 |
| $Zr_2(Al_{0.3}Sb_{0.7})C$ | 220 | 98 | 103 | 243 | 94 | 143 | 73 | 187 | 0.51 | 0.28 | 8.19 |
| $Zr_2AlC$ | 258 | 67 | 63 | 221 | 91 | 124 | 91 | 219 | 0.73 | 0.24 | 16.37 |

**Table 3**
The born criteria and their verification by the elastic constants $Cij$ of the compounds studied herein.

| Born Criteria | $Zr_2(Al_{0.58}Bi_{0.42})C$ | $Zr_2(Al_{0.2}Sn_{0.8})C$ | $Zr_2(Al_{0.3}Sb_{0.7})C$ | $Zr_2AlC$ |
|---|---|---|---|---|
| $C_{11} > 0$ | 242 | 253 | 220 | 258 |
| $(C_{11} - C_{12}) > 0$ | 242-99=143 | 253-108=145 | 220-98=122 | 258-67=191 |
| $C_{44} > 0$ | 129 | 109 | 94 | 91 |
| $(C_{11} + C_{12})C_{33} > 2C_{13}^2$ | 96844>30752 | 105773>24200 | 77274>21218 | 71825>7938 |

The bulk modulus $B$, shear modulus $G$, Young's modulus $Y$, Poisson's ratio $v$ and Pugh ratio $G/B$ are also calculated and given in Table 2. $Y$ and $v$ are calculated from $B$ & $G$ by using the widely used relationships: $Y = 9BG/(3B + G)$ and $v = (3B-Y)/6B$ [32, 33]. The arithmetic average of the Voigt ($B_V$, $G_V$) and the Reuss ($B_R$, $G_R$) bounds are used to estimate the polycrystalline bulk and shear moduli $B$ and $G$ [34]. Furthermore, we have calculated the elastic anisotropy, $A_1$, $A_2$ and $A_3$ and another anisotropic factor, $k_c/k_a$ for the hexagonal crystal defined by the ratio between the linear compressibility coefficients along the $c$- and $a$-axis. There are three independent elastic shear constants for hexagonal crystals; thus, three shear-type anisotropy factors can be defined. $A_1 = \frac{1/6(C_{11}+C_{12}+2C_{33}-4C_{13})}{C_{44}}$, $A_2 = \frac{2C_{44}}{C_{11}-C_{12}}$, $A_3 = A_1 \cdot A_2 = \frac{1/3(C_{11}+C_{12}+2C_{33}-4C_{13})}{C_{11}-C_{12}}$ [35] and $k_c/k_a = (C_{11} + C_{12} - 2C_{13})/(C_{33} - C_{13})$ from the calculated values of the elastic constants.

The value of bulk modulus is found to be increased due to the substitution of Bi/Sn/Sb for Al in $Zr_2AlC$ while the value of shear modulus decreased. The shear modulus describes the material's response to shearing strains *i.e.*, the resistance to change in shape and the bulk modulus describes the material's response to uniform pressure. As we know that the Young's modulus, $Y$ is defined as the ratio of longitudinal stress and longitudinal strain; and is a measure of the stiffness of the solid material. The higher value of $Y$ indicates higher stiffness of the material. Therefore, from Table 2, we see that $Zr_2(Al_{0.58}Bi_{0.42})C$ and $Zr_2(Al_{0.2}Sn_{0.8})C$ are stiffer than both $Zr_2(Al_{0.3}Sb_{0.7})C$ and $Zr_2AlC$.

Poisson's ratio is used often in engineering science, and it relates directly to the failure mode of solids. The critical value of $v$ is 0.26, which identifies a material either as brittle or ductile one [36]. Poisson's ratio $v$ is greater than 0.26 for ductile solids and $v < 0.26$ for brittle materials. Another parameter suggested by Pugh for ductility-brittle transition is the $G/B$ [37] ratio. If $G/B > 0.57$, the material behaves as a brittle, otherwise it is ductile in nature. According to these two conditions, $Zr_2AlC$ is brittle in nature and $Zr_2(Al_{0.58}Bi_{0.42})C$ is in the border line of brittle to ductile transitions. $Zr_2(Al_{0.2}Sn_{0.8})C$ and $Zr_2(Al_{0.3}Sb_{0.7})C$ are ductile in nature. The effect of of Bi/Sn/Sb substitution is noticeable. It is seen from Table 2 that the value of Poisson's ratio is increased and $G/B$ ratio is decreased with Bi/Sn/Sb substitution for Al, indicating a tendency towards increasing ductility.

In order to study the effect of Bi/Sn/Sb substitution on the mechanical hardness of $Zr_2AlC$ and the unsubstituted MAX phase itself, we have calculated the theoretical Vickers hardness using Chen's formula [38], which is expressed as: $H_v = 2(k^2G)^{0.585} - 3$, where, $k = G/B$. Calculated Vickers hardness is given in Table 2. Hardness of a compound is related to both the bulk modulus and the shear modulus. Physically, the bulk modulus only measures isotropic resistance to volume deformation under hydrostatic strain, whereas shear modulus measures the resistance to anisotropic shear strain. Even though it was thought that a direct relation between the bulk modulus and hardness is less obvious than that with the shear modulus [39], the Pugh's ratio ($k = G/B$) clearly shows the contribution of both $B$ and $G$ to the Vickers hardness. From Table 2, it is observed that, the value of $B$ is increased for substituted compounds and the value of $G$ is decreased. The theoretical Vickers hardness of the substituted compounds therefore decreases.

Anisotropy factor of any crystal is quite important in engineering science and also in crystal physics. The degree of anisotropy in the bonding between atoms in different planes is described by the shear anisotropic factors which is calculated and given in Table 4. For an isotropic crystal, the values of $A_1$, $A_2$ and $A_3$ equal to 1, while any value smaller or greater than 1 is a measure of the degree of shear anisotropy possessed by the crystal. It is found that $Zr_2(Al_{0.58}Bi_{0.42})C$, $Zr_2(Al_{0.2}Sn_{0.8})C$ and $Zr_2(Al_{0.3}Sb_{0.7})C$ are more anisotropic than $Zr_2AlC$. In addition, the calculated values of $k_c/k_a$ of $Zr_2(Al_{0.58}Bi_{0.42})C$, $Zr_2(Al_{0.2}Sn_{0.8})C$ and $Zr_2(Al_{0.3}Sb_{0.7})C$ are less than 1, indicating the compressibility along the $c$-axis is smaller than that along the $a$-axis. The reverse is true for $Zr_2AlC$.

Table 4

Anisotropy factors for different shear planes and $k_c/k_a$.

| Compound | $A_1$ | $A_2$ | $A_3$ | $k_c/k_a$ |
|---|---|---|---|---|
| $Zr_2(Al_{0.58}Bi_{0.42})C$ | 0.53 | 1.80 | 0.95 | 0.58 |
| $Zr_2(Al_{0.2}Sn_{0.8})C$ | 0.77 | 1.50 | 1.15 | 0.77 |
| $Zr_2(Al_{0.3}Sb_{0.7})C$ | 0.69 | 1.54 | 1.06 | 0.80 |
| $Zr_2AlC$ | 0.94 | 0.95 | 0.89 | 1.25 |

*3.3 Electronic properties*

The calculated electronic energy band diagrams of the $Zr_2(Al_{0.58}Bi_{0.42})C$, $Zr_2(Al_{0.2}Sn_{0.8})C$, $Zr_2(Al_{0.3}Sb_{0.7})C$ and $Zr_2AlC$ compounds along the high symmetry directions inside the first Brillouin zone are shown in Figs. 2(a), (b), (c) and (d), respectively. The Fermi-level, $E_F$, is set to zero of the energy scale. From the band structures, it is clear that the valence and conduction bands are overlapped considerably and there is no band gap at the $E_F$. As a result all these compounds are expected to show metallic characteristics. There is a strong anisotropic feature along $c$-axis energy dispersion, which can be seen from the reduced level of dispersion along the *H-K* and *M-L* directions. This indicates that these MAX phases should possess anisotropic electrical conductivity.

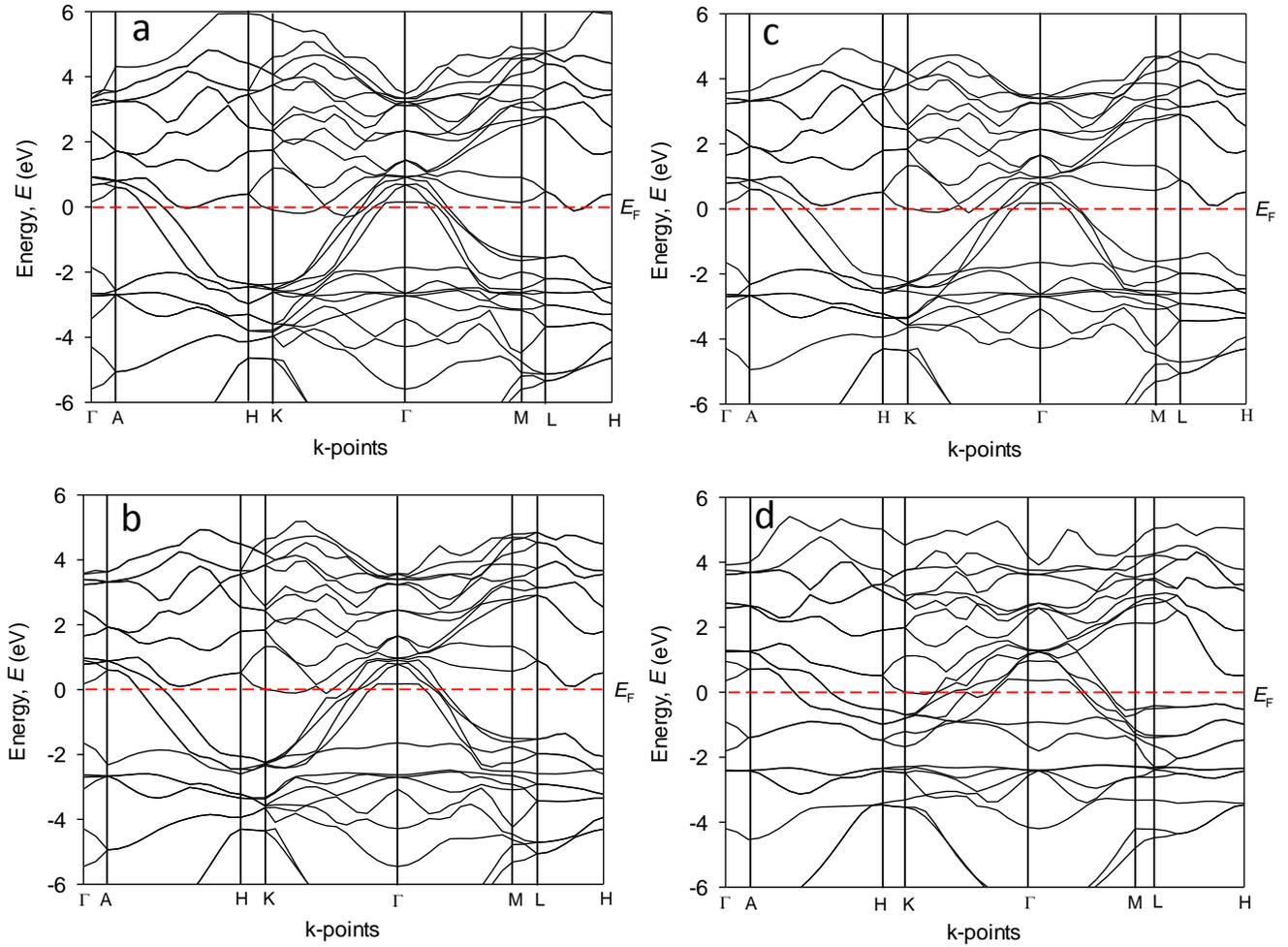

**Fig. 2.** Electronic band structures of (a) $Zr_2(Al_{0.58}Bi_{0.42})C$, (b) $Zr_2(Al_{0.2}Sn_{0.8})C$ (c) $Zr_2(Al_{0.3}Sb_{0.7})C$ and (d) $Zr_2AlC$ MAX compounds.

Total and partial DOS (PDOS) of $Zr_2(Al_{0.58}Bi_{0.42})C$, $Zr_2(Al_{0.2}Sn_{0.8})C$, $Zr_2(Al_{0.3}Sb_{0.7})C$ and $Zr_2AlC$ compounds are represented in Figs. 3(a), (b), (c) and (d), respectively. The value of DOS at the $E_F$, are found to be 3.26, 3.22, 3.65 and 2.72 states/eV for $Zr_2(Al_{0.58}Bi_{0.42})C$, $Zr_2(Al_{0.2}Sn_{0.8})C$, $Zr_2(Al_{0.3}Sb_{0.7})C$ and $Zr_2AlC$, respectively. It is found that the DOS is increased due to Bi/Sn/Sb substitution for Al in $Zr_2AlC$ with the largest value for obtained Sb substitution, indicating the improved electric conduction properties of the substituted compounds. As illustrated in Fig. 3 (a), the Zr $4d$ electrons are the chief contributor to the DOS at the $E_F$ and determine the compounds conduction properties. The contribution from C $2s$, Al $3s$, Zr $5s$ and Bi $6s$ orbitals are not that significant to the DOS at the Fermi level. The valence band is formed due to C $2p$ states with a small admixture of Al $3s/3p$, Zr $4d$, Bi $6s/6p$ orbitals lying between an energy range from -8.5 eV and 0.0 eV (although the data are shown in Figs. 3 from -6.0 eV). The bonding peak at around -2.8 eV indicates to the strong hybridization of Zr 4*d* and C 2*p*/2*s* orbitals in $Zr_2(Al_{0.58}Bi_{0.42})C$. Similar behaviors are observed in case of $Zr_2(Al_{0.2}Sn_{0.8})C$, $Zr_2(Al_{0.3}Sb_{0.7})C$ and $Zr_2AlC$ which are shown in Figs. 3(b), (c) and (d), respectively. The key point of the band structure spectra of the substituted compounds (Figs. 3(a), (b) and (c)) is the appearance of additional band at the Fermi level. The additional band at the Fermi level is attributed to the Bi/Sn/Sb *p* states as shown in Figs. 3. Due to this additional band, the DOS at the Fermi level is enhanced in the substituted MAX nanolaminates.

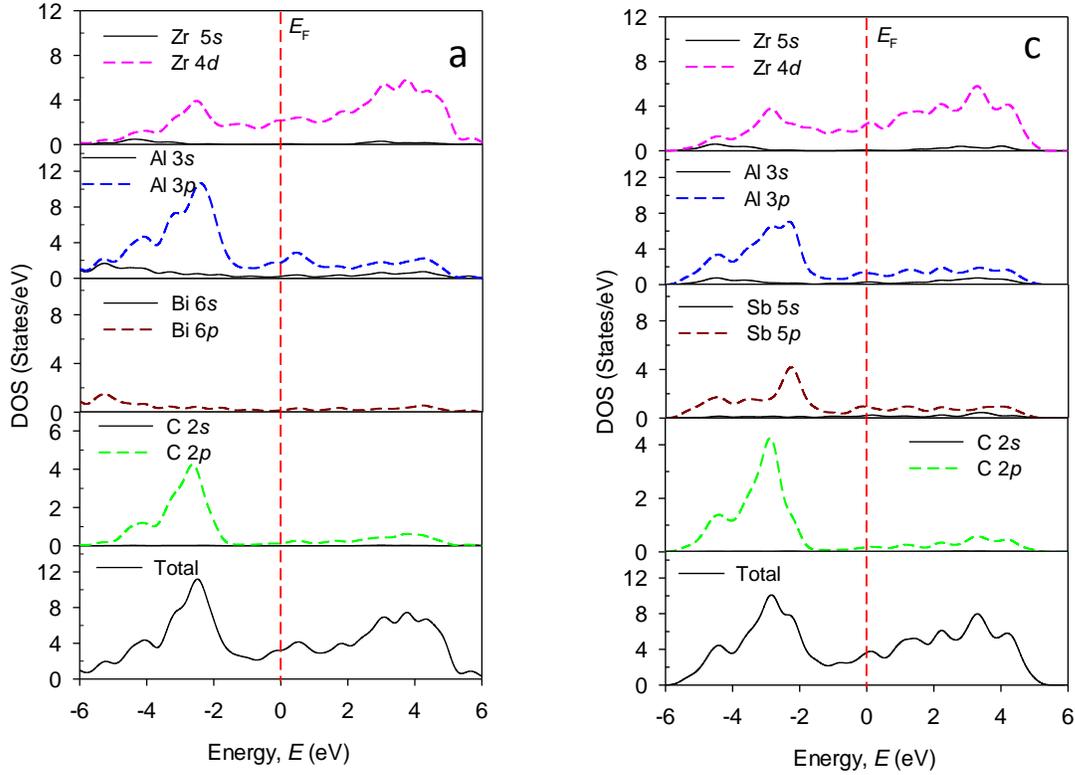

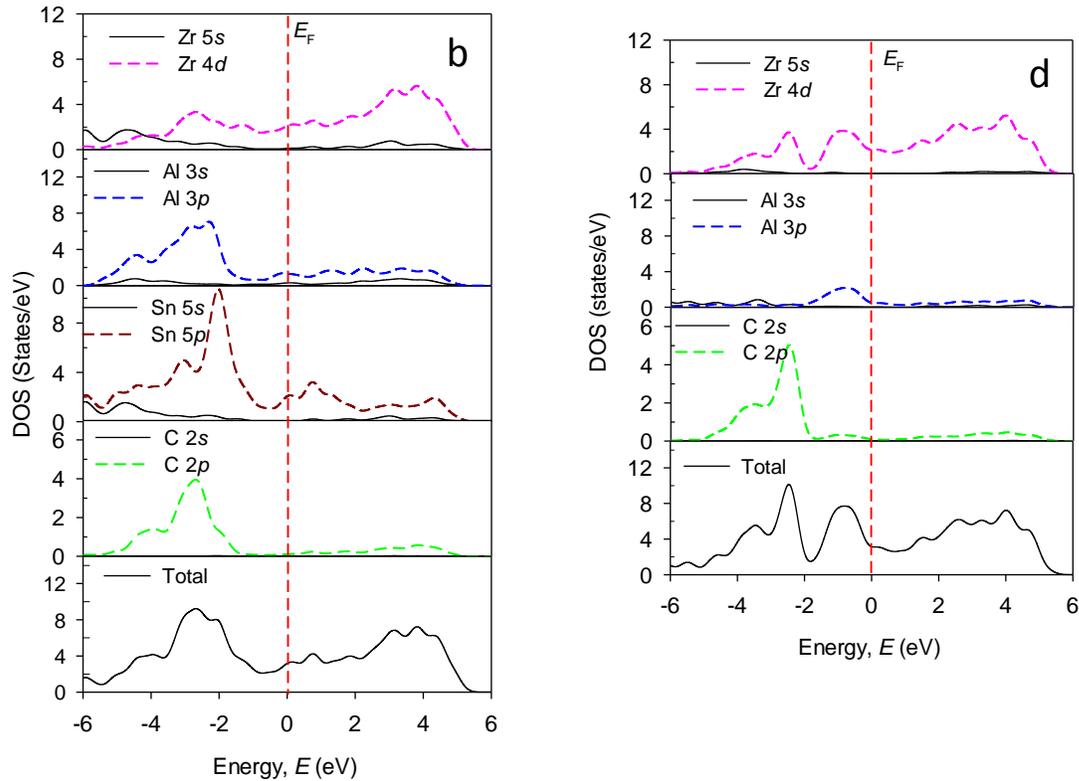

**Fig. 3.** Total and partial density of states for (a) $Zr_2(Al_{0.58}Bi_{0.42})C$, (b) $Zr_2(Al_{0.2}Sn_{0.8})C$ (c) $Zr_2(Al_{0.3}Sb_{0.7})C$ and (d) $Zr_2AlC$ MAX compounds.

Fig. 4 shows the electron charge density map (in the units of $e/Å^3$) along (101) crystallographic plane to characterize the nature of chemical bonding in (a) $Zr_2(Al_{0.58}Bi_{0.42})C$, (b) $Zr_2(Al_{0.2}Sn_{0.8})C$, (c) $Zr_2(Al_{0.3}Sb_{0.7})C$ and (d) $Zr_2AlC$ MAX phases. The red and blue colors in the adjacent scale to the map indicate the low and high intensity of electron charge density, respectively. The high intensity of electron charge density indicates the strong accumulation of electronic charges and vice versa. Some significant variations in the charge density distribution were observed with the substitution of Bi/Sn/Sb for Al. The size of tubular symmetric shape was increased with Bi/Sn/Sb substitution for Al. As shown in Fig. cylindrical/tubular symmetric shapes of charge density between Zr-Al, Zr-Al/Bi, Zr-Al/Sn and Zr-Al/Sb could be identified as covalent bonding, while spherical symmetric shapes between Zr-C atoms indicate ionic bonding. The strong hybridization of Zr $4d$ and C $2p/2s$ are responsible for the formation of ionic bonding. It is also evident that covalent bonding Al-Al exists between Al-atoms. The iconicity/covalency of the bond increase with the increase of electron density.

The present results are in well consistent with the findings of electronic properties (DOS) of the compound.

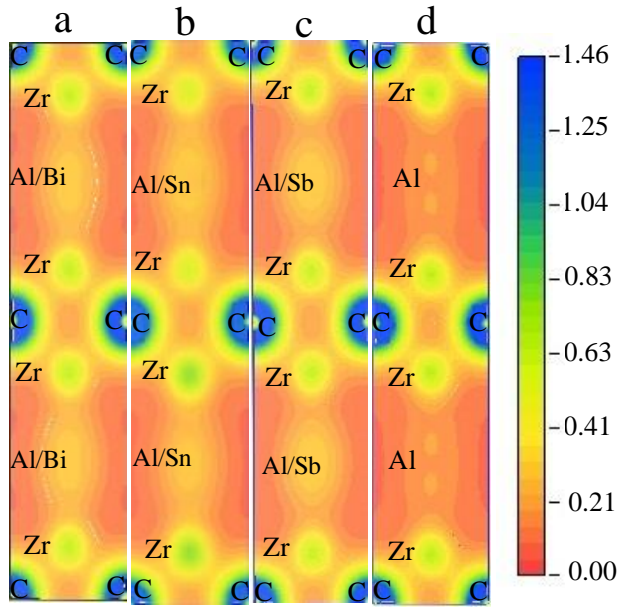

Fig. 4 Distribution of charge density mapping images of (a) $Zr_2(Al_{0.58}Bi_{0.42})C$, (b) $Zr_2(Al_{0.2}Sn_{0.8})C$, (c) $Zr_2(Al_{0.3}Sb_{0.7})C$ and (d) $Zr_2AlC$ MAX phases.

## 4. Conclusion

We have calculated the structural, elastic, Vickers hardness, and electronic band structure properties of very recently discovered $Zr_2(Al_{0.58}Bi_{0.42})C$, $Zr_2(Al_{0.2}Sn_{0.8})C$, $Zr_2(Al_{0.3}Sb_{0.7})C$, and $Zr_2AlC$ MAX phases for the first time using first principles density functional formalism. The optimized lattice parameters agree reasonably well with their experimental counterparts. Various technologically important elastic constants and moduli are estimated for the first time. The bulk modulus increases significantly in the substituted compounds in comparison to $Zr_2AlC$. The Young's modulus, on the other hand remains almost unchanged, except in $Zr_2(Al_{0.3}Sb_{0.7})C$, where a marked decrease is noted. The elastic constants satisfy the mechanical stability criteria for all the compounds under consideration. Mechanical anisotropy factor increases due to Sn/Sb/Bi substitutions; $Zr_2(Al_{0.58}Bi_{0.42})C$ being the most anisotropic. The increase of bulk modulus and decrease of shear modulus result to substantially lower Vickers hardness in the Bi/Sn/Sb substituted samples with respect to the $Zr_2AlC$. The electronic band structures show that all these MAX compounds are metallic in nature. The electronic density of states at the Fermi level in the substituted phases are somewhat enhanced due to the contributions from the Bi/Sn/Sb $5p$ orbitals. Substantial electronic anisotropy is expected in these compounds due to non-dispersive nature of the band structure along $c$-direction. We hope that this theoretical investigation of mechanical and electronic properties of technologically promising and recently discovered $Zr_2(Al_{0.58}Bi_{0.42})C$, $Zr_2(Al_{0.2}Sn_{0.8})C$, $Zr_2(Al_{0.3}Sb_{0.7})C$, and $Zr_2AlC$ MAX nanolaminates will stimulate further experimental and theoretical studies in near future.

**Compliance with Ethical Standards**

The authors declare that they have no conflict of interest.